\documentclass[preprint,aps,12pt,showpacs,nofootinbib,tightenlines]{revtex4}
\usepackage{amsmath}
\usepackage{amssymb}
\usepackage{epsfig}
\usepackage{graphicx}
\usepackage{subfigure}
\usepackage{slashed}
\usepackage{multirow}
\usepackage{color}
\usepackage[]{natbib}
\usepackage[colorlinks, citecolor=blue, urlcolor=blue, linkcolor=blue]{hyperref}
\newcommand{\met}{\not\!\!\!E_{T}}

\begin{document}

\def\pslash{\rlap{\hspace{0.02cm}/}{p}}
\def\eslash{\rlap{\hspace{0.02cm}/}{e}}
\title {Search for the singlet vector-like top quark in the $T\to tZ$ channel with $Z\to \nu\bar{\nu}$ at hadron colliders}
\author{Lin Han $^{1,2}$}\email{hanlin@xxmu.edu.cn}
\author{Shiyu Wang $^2$}
\author{Liangliang Shang $^2$}
\author{Bingfang Yang$^2$}
\affiliation{$^1$ School of Medical Engineering, Xinxiang Medical University, Xinxiang 453003, China\\
	$^2$ School of Physics, Henan Normal
	University, Xinxiang 453007, China
	\vspace*{1.5cm}  }

\begin{abstract}

Based on a simplified model including a singlet vector-like top quark $T$ with charge $|Q|=2/3$, we analyze the prospects of observing $T$ via the single $T$ production in the $tZ$ channel with $Z$ decaying to neutrinos at the hadron-hadron colliders. This simplified model only includes two free parameters, the coupling constant $g^*$
and the $T$ quark mass $m_T$. To investigate the observability of the single $T$ production, we perform a detailed background analysis and detector simulation for the collision energies 14~TeV, 27~TeV, and 100~TeV. We scan the $g^*-m_T$ parameter space and show the exclusion and discovery capabilities on the $T$ quark with the highest integrated luminosity designed at these colliders. Moreover, the limits from the narrow-width approximation and electroweak precision observables are considered.
\end{abstract}
	
\pacs{14.65.Jk,13.66.Hk,12.60.-i}
\maketitle
\newpage
\section{ INTRODUCTION}

The discovery of a 125~GeV Higgs boson by the ATLAS and CMS collaborations  at  the  Large  Hadron  Collider  (LHC) completed the last piece of the Standard Model (SM) of particle physics. Meanwhile, the Higgs data have excluded the possibility of additional SM-like chiral fermions. In contrast, the vector-like quarks (VLQs) \cite{VLQs} are consistent with existing Higgs data since they do not receive their masses from Yukawa couplings to a Higgs doublet.  The VLQs are color-triplet spin-1/2  fermions,  and  the  left- and right-handed components transform with the same properties under the SM electroweak symmetry group. In some new physics models\cite{VLQ model1, VLQ model2,VLQ model3,VLQ model4}, such as
Little Higgs\cite{Arkani-Hamed:2002ikv, M. Schmaltz,Hsin-Chia Cheng,Spencer Chang} and Composite Higgs\cite{D.B. Kaplan, Low:2015nqa,Jing Shu} models, the vector-like top quark (VLT) is often introduced to alleviate the gauge hierarchy problem since the VLT is arranged to cancel the one-loop quadratic divergence of the Higgs mass parameter induced by the top quark. 

In experiment, the VLT with masses
below 1 TeV are mainly pair-produced via the strong interaction at the LHC. The single production of VLT via the electroweak interaction is also important and may have a larger cross section for the VLT with masses above 1 TeV due to weaker phase-space suppression. Recently, the searches for VLT have been performed in single and pair-produced modes at the LHC with $\sqrt{s}$ = 13~TeV. Here, we only focus on the search for the singlet VLT. 
\begin{itemize}
	\item In the single-produced process: 
(1) For the $T\to tH$ or $T\to tZ$ channel, the ATLAS collaboration performed a search and presented that the singlet $T$ quark mass below 1.8(1.6) TeV was excluded for the universal coupling strength $\kappa$ values above 0.5(0.41) corresponding to 139 fb$^{-1}$\cite{T-tz-th1}. Meanwhile, the CMS collaboration performed a search for the channel $T$ → $tZ$ and presented that the singlet $T$ quark mass below 1.4 TeV was excluded for a resonance of fractional width in the range 10\% to 30\% corresponding to 136 fb$^{-1}$\cite{T-tz-th2}. (2) For the $T$ → $Wb$ channel, the search performed by the ATLAS collaboration has set the upper limits on the singlet $T$ quark of mass 800 GeV for the mixing angle $|\sin\theta_L| = 0.18$ corresponding to 36.1 fb$^{-1}$\cite{T-wb}. 
\item In the pair-produced process: 
(1) For various decay channels ($T\to Wb/tZ/tH$), the singlet $T$ quark is
excluded for masses below 1.31 TeV corresponding to 36.1 fb$^{-1}$\cite{pair-all}. (2)  For the $T$ → $tZ$ channel, the limits on the singlet $T$ are set at $m_{T}$ > 1.27 TeV corresponding to 139 fb$^{-1}$\cite{pair-tz}. 
\end{itemize}

In phenomenology, the studies of VLT have been performed extensively in general decay modes ($Wb, tZ, tH$)\cite{Matsedonskyi:2014mna,Andeen:2013zca,Vignaroli:2012sf,Vignaroli:2012nf, DeSimone:2012fs,T-NPB-Han,T-wb-prd-hou,T-wb-prd-wang, Buckley:2020wzk, Belyaev} or some exotic channels\cite{Senol:2011nm,Alwall:2010jc,Liu:2015kmo,T-prd-Wang,T-prd-zhou,T-FC}. Especially, the VLT can be probed at the LHC by the same-sign dilepton signature\cite{TT-prd-liu, T-ctp-zhou,tt-prd-jjcao}. In Refs.\cite{T-tz-liu,T-tz-yang}, the authors have studied the single production of singlet VLT via $T\to tZ(Z\to ll)$ at the high luminosity (HL)-LHC with 14 TeV \cite{HL-LHC}, the high energy (HE)-LHC with 27 TeV \cite{HE-LHC} and the Future Circular Hadron Collider (FCC-hh) with 100 TeV\cite{FCC-hh}. In this study, we investigate the single production of the singlet VLT decaying into $tZ(Z\to \nu\bar{\nu})$ at high energy hadron colliders since searches of the VLT on $Z\to ll$ and $Z\to \nu\bar{\nu}$ channels via experiments are also independently carried out.
For the high energy hadron colliders, the single production of the VLT decaying into $tZ$, followed by the $Z$ boson decaying into neutrinos, results in a mono-top signature, which is evidently different from Dark Matter (DM) production as can be seen in the following section. This channel search has been performed by the LHC experiment\cite{T-tz-th2}, and we expect this study to provide a theoretical reference for future analysis and search at the LHC and future hadron colliders.

The paper is organized as follows. In Sec.II, we briefly review the Lagrangian of the singlet VLT and discuss the limits on the model parameters from the current experiments. In Sec.III, we describe the event generation and detector simulation of the signal and backgrounds at hadron colliders. In Sec.IV, we show the observability of the signal at the HL-LHC, HE-LHC and FCC-hh. Finally, we summarize our results in Sec.V.


\section{SINGLET VLT MODEL}

The Lagrangian of the $SU(2)$ singlet VLT with couplings only to the third generation of SM quarks can be expressed as \cite{Buchkremer:2013bha}.
\begin{eqnarray}
\mathcal{L}_T = \frac{gg^{*}}{2}\{\frac{1}{\sqrt{2}}[\bar{T}_LW^{+}_{\mu}\gamma^{\mu}b_L]
+\frac{1}{2\cos{\theta_W}}[\bar{T}_LZ_{\mu}\gamma^{\mu}t_L] \cr
-\frac{m_T}{2m_W}[\bar{T}_RHt_L] -
\frac{m_t}{2m_W}[\bar{T}_LHt_R]\} + h.c.
\end{eqnarray}
where $g^{*}$ is the coupling strength of the $T$ quark only entering the single production to SM quarks.
$g$ is the $SU(2)_L$ gauge coupling constant, and $\theta_W$ is the Weinberg angle.
In this simplified model, the $T$ quark mass $ m_T $ and the coupling strength $g^{*}$ are the only two free parameters.  

For the coupling coefficient, different symbols are used in different studies\cite{ATLAS:2016ovj,Buchkremer:2013bha};  the relationship of these symbols can be deduced as follows:
 \begin{eqnarray}
 \label{huansuanguanxi}
g^{*}=\sqrt{2} \kappa_{T}=2 \sin \theta_{L}
 \end{eqnarray}
As mentioned above, the limit on $g^{*}$ from the LHC direct searches can be conservatively set to a range, $ g^{*} $ $ \le $  0.5, which is weaker than the limit from the electroweak precision observables (EWPOs)\cite{Aguilar-Saavedra:2013qpa}. Here, we consider the EWPO limit by the oblique parameters $S, T, U$\cite{STU1,STU2} and take the $S, T, U$ experimental values as\cite{PDG2020}
\begin{eqnarray}
S=-0.01\pm 0.10,~~
T=0.03\pm0.12,~~
U=0.02\pm0.11.
\end{eqnarray}
There is a strong correlation (0.92) between the
$S$ and $T$ parameters. The $U$ parameter is -0.8 (-0.93) anti-correlated with $S (T)$. We adopt the methods in Refs.\cite{cao-wmass,Crivellin-wmass} to calculate this limits and show them in the following figures of numerical results.

\section{EVENT GENERATION}

In hadron-hadron collisions, bottom-quarks arise at leading order (LO) in $\alpha_s$ through the splitting of a gluon. In the four flavor scheme (4FS), one does not consider $b$-quarks as partons in the proton, this splitting is described by fixed-order perturbation theory, and includes the full dependence on the transverse momentum $p_T$ of the $b$-quark and its mass.  In the five flavor scheme (5FS), the splitting arises by solving the DGLAP evolution equations with five massless quark flavours\cite{DGLAP}. In recent direct searches for the VLQs at the LHC, the 4FS was implemented in most experiments\cite{T-tz-th1, T-tz-th2, T-wb, 1909.04721, PAS B2G-16-001, PAS B2G-17-007, ATLAS-tz-nunu}; therefore, we perform the theoretical simulation under this scheme.

\begin{figure}[htbp]
	\scalebox{0.6}{\epsfig{file=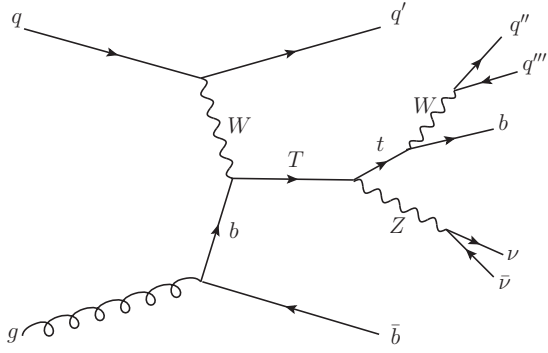}}\vspace{-0.5cm}
	\caption{Feynman diagram for the production of single $T$ quark decaying to a top quark and a $Z$ boson with $Z\to \nu\bar{\nu}$ at the $pp$ collider.}
	\label{fig:Feynman}
\end{figure}

We explore the observability of the signal through the following process
$$ qg \rightarrow q'\ T (\rightarrow tZ)\bar{b}\rightarrow q'\ t (\rightarrow bjj)\ Z (\rightarrow \nu \bar{\nu})\bar{b} \rightarrow 3j+2b+\met $$
and the related Feynman diagram is shown in Fig.\ref{fig:Feynman}. 
We can see that the heavy $T$ quark
produced in $pp$ collisions via $Wb$ fusion and the signal events contain three light jets, two $b$ jets, and missing energy $\met$. One important difference between DM production and VLT production is the presence of additional quarks in the single production of $T$ quarks, which will lead to at least one jet being detected at a small angle relative to the beam line. Similar to the DM case, the topology of the VLT signal has a distinctive signature, characterised by the presence of a top-quark and missing transverse momentum arising from the $Z\to \nu\bar{\nu}$ decay. In the detector simulation, we choose $N_j \ge 3 , N_b \ge 1 , N_l = 0 $ as the trigger condition.  According to these signal characters, the main SM backgrounds are $Z$+jets, $t$$\bar{t}$, $t$$\bar{t}$$V(V=W,Z)$, $VV(V=W,Z)$, $tX(X=j,W,b)$ and $t\bar{b}Zj$. To consider the background more fully, we make the top quark and $W$ boson decay into all final states in backgrounds except for the irreducible background $t\bar{b}Zj$. For clarity, we summarize the production processes and decay modes of the backgrounds in Table.\ref{tab:process}. In our calculations, the signal conjugate process $ qg \rightarrow q'\ \bar{T}b$ and the background conjugate processes listed in Table.\ref{tab:process} have been included.

\begin{table}[]
	\centering
	\caption{Production processes and decay modes of the backgrounds, where the conjugate processes are shown in parentheses. For simplicity, we do not present the decay modes of the conjugate processes.}
	\label{efficiency-1}
	\newcommand{\tabincell}[2]{\begin{tabular}{@{}#1@{}}#2\end{tabular}}
	\resizebox{0.8\textwidth}{!}{
		\begin{tabular}{c|c|c||c|c|c}
			\hline \hline
			&\begin{tabular}[c]{@{}c@{}}Backgrounds\end{tabular} & \begin{tabular}[c]{@{}c@{}} Decay mode \end{tabular} & &\begin{tabular}[c]{@{}c@{}}Backgrounds\end{tabular} & \begin{tabular}[c]{@{}c@{}} Decay mode \end{tabular} \\ 
			\hline			
			\multirow{3}{*}{\tabincell {c}{Single\\top}}&$pp\rightarrow tj(\bar{t}j)$  &
			\begin{tabular}[c]{@{}c@{}} $t\rightarrow all$
			\end{tabular}
			&\multirow{3}{*}{Diboson} &$pp\rightarrow W^{+}W^{-} $  &
			\begin{tabular}[c]{@{}c@{}} $W^{+} \rightarrow {all}$, $W^{-} \rightarrow all$
			\end{tabular}\\ 
			\cline{2-3}
			\cline{5-6}
			&$pp\rightarrow tW^{-}(\bar{t}W^{+})$  &
			\begin{tabular}[c]{@{}c@{}} $t\rightarrow all$, $W^{-} \rightarrow {all}$
			\end{tabular}
			& &$pp\rightarrow W^{+}Z(W^{-}Z)$   &
			\begin{tabular}[c]{@{}c@{}} $W^{+} \rightarrow {all}$, $Z \rightarrow \bar \nu_{l}\nu_{l}$
			\end{tabular}\\ 
			\cline{2-3}
			\cline{5-6}
			&$pp\rightarrow t\bar{b}(\bar{t}b)$  &
			\begin{tabular}[c]{@{}c@{}} $t\rightarrow all$
			\end{tabular}& &$pp\rightarrow ZZ$  &
			\begin{tabular}[c]{@{}c@{}} $Z \rightarrow \bar \nu_{l}\nu_{l}$,$Z \rightarrow all$
			\end{tabular}\\ 
			\cline{1-6}
			
			\multirow{3}{*}{\tabincell {c}{Top\\pair}} &$pp\rightarrow t\bar{t}$   &
			\begin{tabular}[c]{@{}c@{}}$t\rightarrow {all}$, $\bar{t} \rightarrow {all}$ 
			\end{tabular} &$Z$+jets &$pp\rightarrow Zjjj$  &
			\begin{tabular}[c]{@{}c@{}} $Z \rightarrow \bar \nu_{l}\nu_{l}$
			\end{tabular}\\
			\cline{2-3}
			\cline{4-6}
			&$pp\rightarrow t\bar{t}W^{+}(t\bar{t}W^{-})$   &
			\begin{tabular}[c]{@{}c@{}}$t\rightarrow {all}$, $\bar{t}
				\rightarrow {all}$, $W^{+} \rightarrow {all}$
			\end{tabular}&Other &$pp\rightarrow t\bar{b}Zj(\bar{t}bZj)$   &
			\begin{tabular}[c]{@{}c@{}}  $t\rightarrow bjj$, $Z \rightarrow \bar \nu_{l}\nu_{l}$
			\end{tabular}\\ 
			\cline{2-3} 
			\cline{4-6}
			&$pp\rightarrow t\bar{t}Z$   &
			\begin{tabular}[c]{@{}c@{}}$t\rightarrow {all}$, $\bar{t}
				\rightarrow {all}$, $Z \rightarrow {\bar \nu_{l}\nu_{l}}$
			\end{tabular}\\
			\hline  \hline 
	\end{tabular}}
	\label{tab:process}
\end{table}

For the signal and backgrounds, we calculate the LO cross sections and generate the parton-level events by MadGraph5-aMC@NLO\cite{MG5_aMC_v2_6_6}, where the CTEQ6\_L\cite{Pumplin:2005rh} is used as the parton distribution function (PDF), and the renormalization and factorization scales are set dynamically by default. The numerical values of the input SM parameters are taken as follows\cite{PDG2020}:
\begin{align}
\nonumber m_t = 172.76{\rm ~GeV},\quad &m_{Z} =91.1876 {\rm ~GeV}, \quad m_h =125.10 {\rm ~GeV}, \\
\nonumber \sin^{2}\theta_W& = 0.231,\quad \alpha(m_Z) = 1/128.
\end{align}
The basic cuts for the signal and backgrounds are choosen as follows:
\begin{eqnarray}\label{basic}
\nonumber  \Delta R(x,y) >  0.4\ &,&\quad  x,y =  \ell, \ j , \ b \\
\nonumber  p_{T}^\ell > 25 \ \text{GeV}&,&\quad  |\eta_\ell|<2.5  \\
\nonumber  p_{T}^j > 25 \ \text{GeV}&,&\quad  |\eta_j|<5.0  \\
\nonumber  p_{T}^b > 25 \ \text{GeV}&,&\quad  |\eta_b|<5.0
\end{eqnarray}

We renormalize the LO cross sections of the backgrounds to the next-leading-order(NLO) or the next-next-leading-order(NNLO) cross sections by multiplying by a $K$ factor. We ignore the differences of the $K$ factor at the HL-LHC, HE-LHC, and FCC-hh and take the values listed in Table.\ref{tab:kfactor} for the different processes. 

\begin{table}[!htb]
	\centering
	\caption{ $K$-factors of the QCD corrections for the background processes.}
	\label{tab:kfactor}
	\resizebox{0.8\textwidth}{!}{
		\begin{tabular}{c|c|c|c|c|c|c|c|c|c|c|c}
			\hline \hline
			&$Z$+jets & \multicolumn{3}{c|}{Top-pair}& \multicolumn{3}{c|} { Single top } & \multicolumn{3}{c|} { Diboson } & \multicolumn{1}{c} { Other } \\
			\hline
			Processes & $Zjjj$ & $~~~~t\bar{t}~~~~$ & $~~~~t\bar{t}W~~~~$ &  $~~~~t\bar{t}Z~~~~$ & $~~~~tj~~~~$  & $~~~tW^{-}~~~$ & $~~~~t\bar{b}~~~~$ & $~WW~$ & $~WZ~$ & $~ZZ~$ & $~~~t\bar{b}Zj~~~$ \\
			\hline
			K-factor & 1.2\cite{MG5_aMC_v2_6_6} &1.8~\cite{Czakon:2012zr}
			&1.2~\cite{Kidonakis:2018ncr}\cite{Campbell:2012dh}
			&1.3~\cite{Kidonakis:2018ncr}\cite{Campbell:2012dh} &1.4~\cite{Kidonakis:2018ncr}\cite{Boos:2012vm} &1.6~\cite{Kidonakis:2018ncr}\cite{Boos:2012vm} &1.9~\cite{Kidonakis:2018ncr}\cite{Boos:2012vm} &1.6~\cite{Campbell:1999ah} &1.7~\cite{Campbell:1999ah} &1.3~\cite{Campbell:1999ah} &1.1~ \\
			\hline \hline
	\end{tabular}}
	
\end{table}

The $T$ quark is significantly heavy, and therefore, the daughter top quark and $Z$ boson are boosted highly. In this case, the C-A reconstruction algorithm \cite{CMS:2009lxa} is a better choice compared with the conventional anti-$kt$ algorithm\cite{Cacciari:2005hq}. Thus, we adopt the C-A algorithm to reconstruct the signal and backgrounds.

We transmit these parton-level events to Pythia 8\cite{PYTHIA} for the parton shower. Then, we perform a fast detector simulation by Delphes 3.14\cite{DELPHES}, where the CMS cards of
the LHC, HE-LHC, and FCC-hh are adopted. We use Fastjet\cite{FastJet} to cluster jets with the C-A algorithm, where the distance parameter is fixed at $\Delta R = 1.5$. Finally, we use MadAnalysis 5\cite{MadAnalysis} to
perform the event analysis. During program operation, we apply the package EasyScan\_HEP\cite{Easyscan} to connect these programs and scan the parameter space. We evaluate the expected signal significance by the Poisson formula\cite{ss}:
\begin{equation}\label{eq:ss}
SS = \sqrt{2 L [(\sigma_S + \sigma_B)\ln(1+\frac{\sigma_S}{\sigma_B})-\sigma_S]}
\end{equation}
where $ L $ denotes the integrated luminosity, $ \sigma_{S}$ and $\sigma_{B}$ denote the cross sections after all cuts for signal and backgrounds, respectively. 
\section{observability}
In this section, we analyze the observability and calculate the statistical significance of the signal at the HL-LHC, HE-LHC, and FCC-hh colliders.

\subsection  { $\sqrt{s}$ = 14~TeV }

\begin{figure}[!htb]
	\begin{center}
		\includegraphics[width=0.5\linewidth]{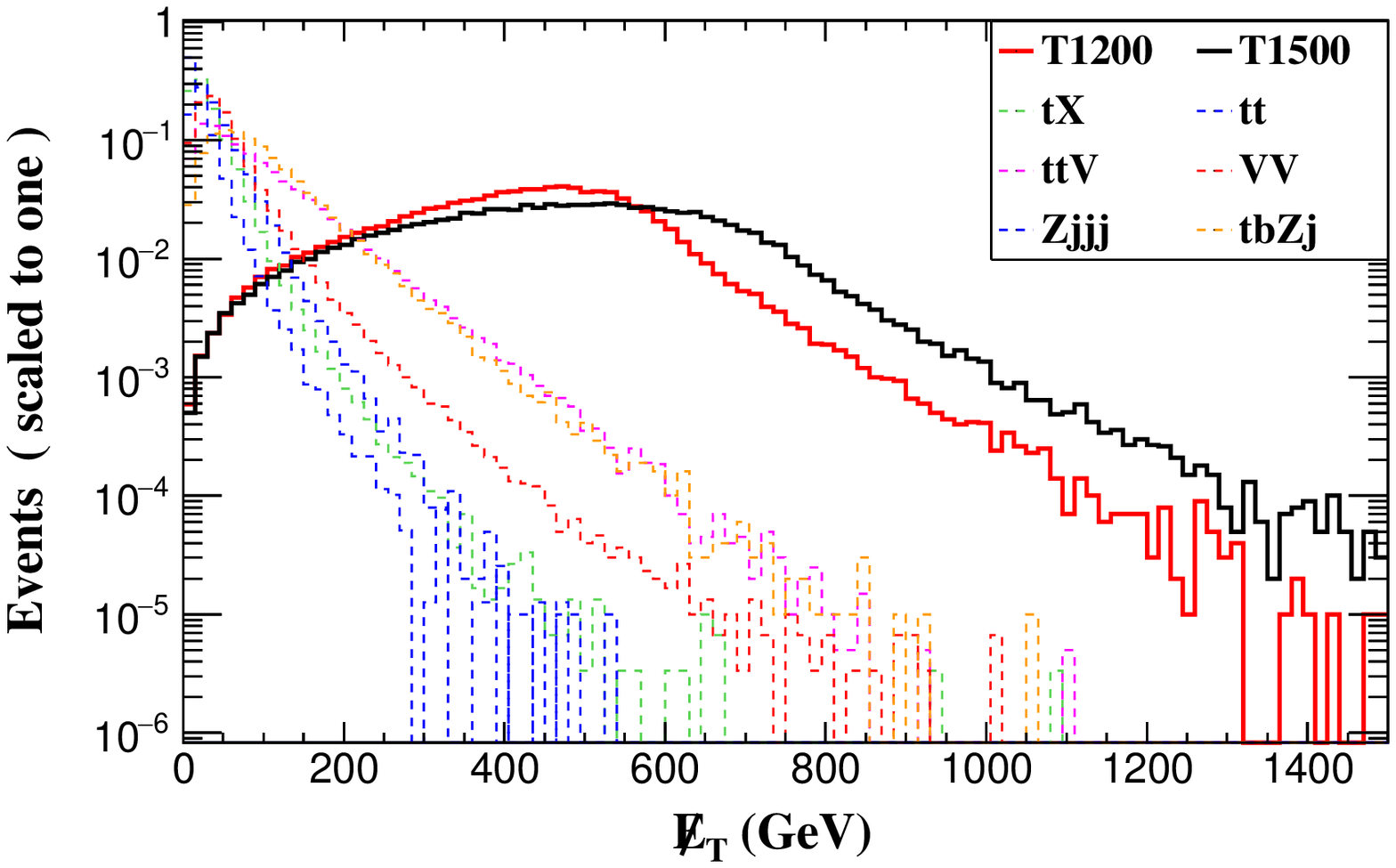}\hspace{-0.5cm}
		\includegraphics[width=0.5\linewidth]{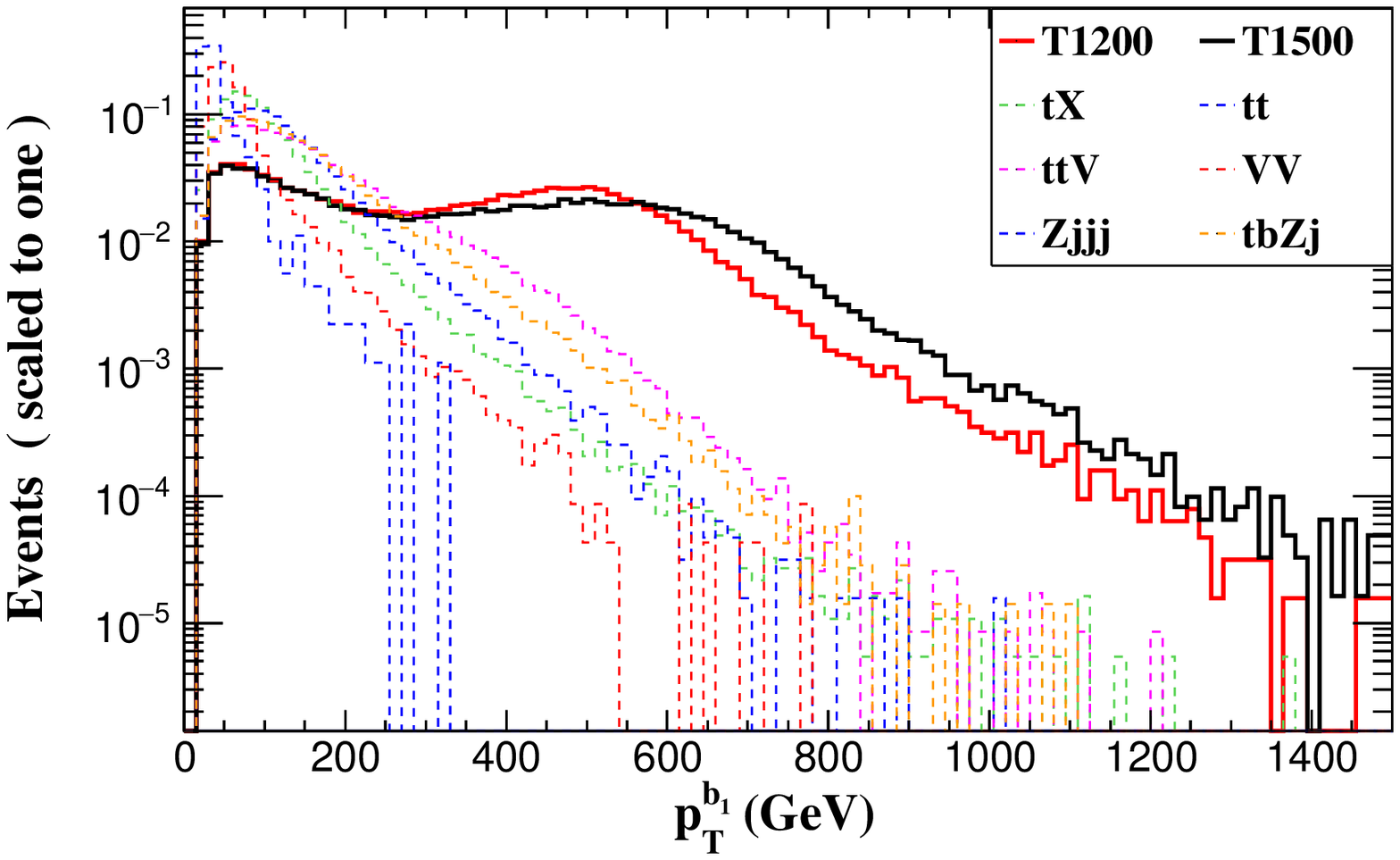}\\
		\includegraphics[width=0.5\linewidth]{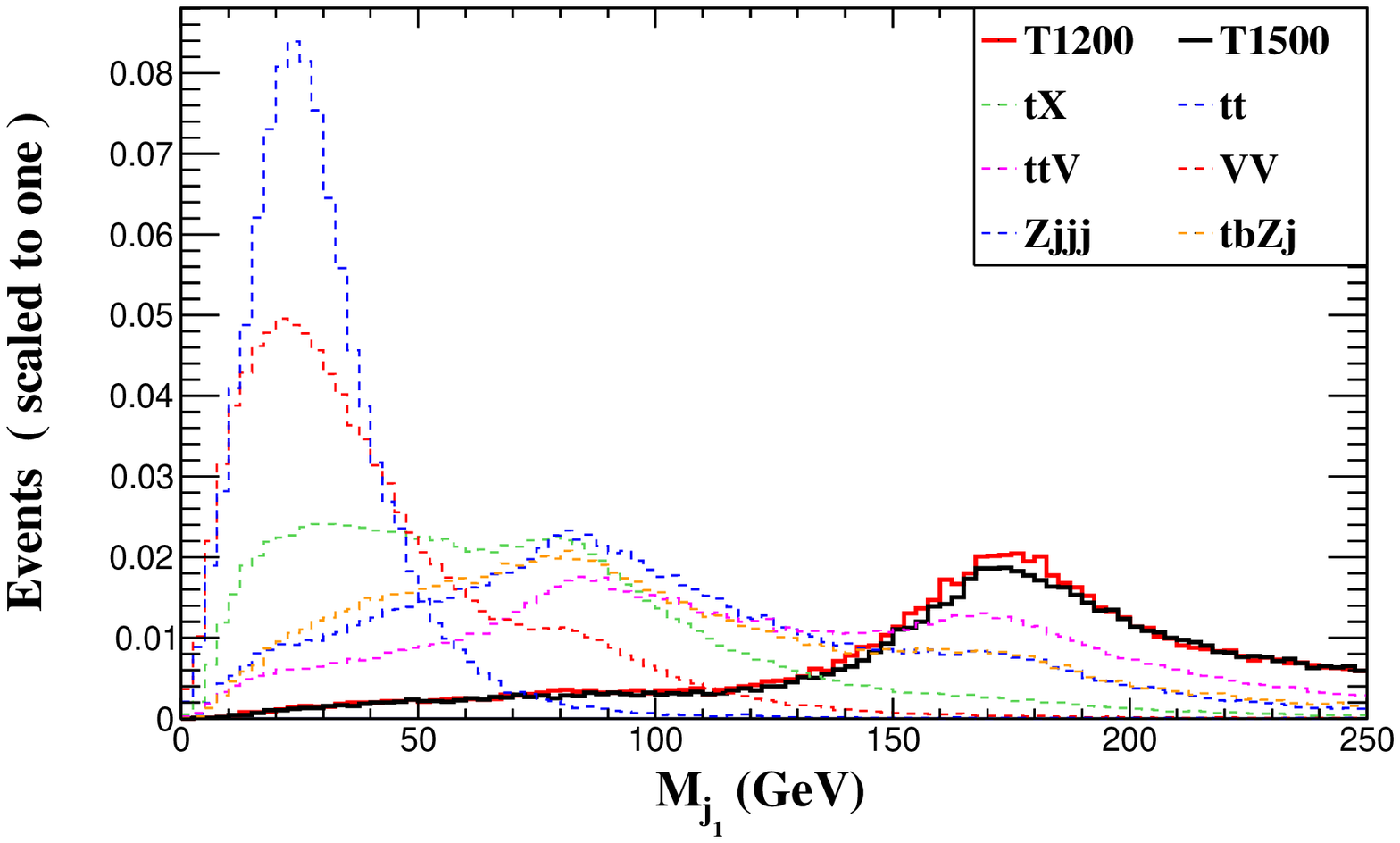}\\
		\caption{Normalized distributions of $\slashed{E}_T$, $p_T^{b_1}$, and $M_{j_1}$ for the two signal benchmark points (T1200 and T1500) and backgrounds with $g^*=0.2$ at the 14~TeV LHC.
			}
		\label{fig:14TeV_distribution}
	\end{center}
\end{figure}

For the signal, the $Z$ boson is boosted highly so that the large missing energy $\slashed{E}_T$ from a pair of neutrinos is expected. Meanwhile, the decay products of the top quark are collimated and captured in a large-radius (large-$R$) jet. Moreover, the leading $b$ jet that comes from the $T$ decay in the signal will have large transverse momentum due to the boosted effect.  Based on the above analysis, we choose the missing energy $\slashed{E}_T> 450$ GeV,  transverse momentum $p_{T}^{b_{1}}> 100$ GeV, and leading large-$R$ jet-mass 160 GeV$<M_{j_1}<$190 GeV as the selection criteria. We show these distributions at the 14 TeV LHC in Fig.\ref{fig:14TeV_distribution}, where $m_T$ = 1500 1200~GeV (labeled as T1500 and T1200) are chosen as two benchmark points.  We summarize the cut flows of the signal and backgrounds in Table.\ref{tab:14TeV_cutflows} and can observe that the largest backgrounds come from $tX$, $t\bar{t}$, and $Zjjj$ before cuts. After the selected cuts, all the backgrounds can be suppressed efficiently, and the total cut efficiency of the signal can reach 6.0\% (5.4\%) for the benchmark point T1500 (T1200). It is worth noting that the backgrounds $tX, t\bar{t}, VV, Zjjj$ are negligibly small after the selected cuts. We have generated events on the order of $10^{6}$ or more for these backgrounds and found that the remaining events are still negligible. Considering the statistical fluctuation, we take $10^{-7}$ as an optimistic estimate of their cut efficiencies. To investigate the exclusion and discovery capabilities on the VLT at the 14 TeV LHC, we scan the parameter space $g^*\in[0.05, 0.5]$ and $m_{T}\in[1000\text{GeV}, 2000\text{GeV}]$, where the average value 5.7\% of these two signal efficiencies is imposed on the entire parameter space.

\begin{table}[!htb]
	\centering
	\caption{Cut flows of the signal (T1200 and T1500) and backgrounds for $g^*=0.2$ at the LHC. }
	\label{tab:14TeV_cutflows}
	\resizebox{1\textwidth}{!}{
		\begin{tabular}{cccccccc}
			\hline
			Cuts&\multicolumn{1}{c}{Signal (fb)}&\multicolumn{6}{c}{Backgrounds (fb)}\\
			\cline{3-8}
			& T1500(T1200)   &   $tX$    &   $t\bar{t}$  &   $ttV$  &    $VV$   &    $Zjjj$   &   $tbZj$  \\ 
			\hline
			$\sigma$ (Before cut)    &   $1.20\times10^{-1}$($4.10\times10^{-1}$)   & $1.03\times10^{5}$  & $2.17\times10^{5}$  &  $7.98\times10^{1}$  & $1.25\times10^{4}$  & $1.91\times10^{5}$  &  $1.66\times10^{1}$  \\
			\hline
			Trigger  & $7.04\times10^{-2}$($2.49\times10^{-1}$)  & $4.46\times10^{4}$   & $8.28\times10^{4}$  &  $2.86\times10^{1}$  & $5.0\times10^{2}$   &  $1.21\times10^{3}$  & $1.12\times10^{1}$   
			\\
			$\met > 450$  GeV       & $3.40\times10^{-2}$($1.06\times10^{-1}$)  & $2.05\times10^{1}$     &  $2.17\times10^{-2}$    &  $1.20\times10^{-1}$   &   $4.15\times10^{-1}$   &  $1.91\times10^{-2}$   &  $3.70\times10^{-2}$ 
			\\
			$p_{T}^{b_{1}} > 100$ GeV         & $3.12\times10^{-2}$($8.67\times10^{-2}$)  & $1.37\times10^{1}$  &  $2.17\times10^{-2}$   &  $9.04\times10^{-2}$  &  $2.49\times10^{-1}$    &  $1.91\times10^{-2}$   &  $2.94\times10^{-2}$    
			\\
			160 GeV$ < M_{j_{1}} < 190$ GeV       & $7.24\times10^{-3}$($2.2\times10^{-2}$)  & $1.03\times10^{-2}$   &  $2.17\times10^{-2}$   &  $1.50\times10^{-2}$  &  $1.25\times10^{-3}$     &  $1.91\times10^{-2}$   &  $4.64\times10^{-3}$    
			\\
			\hline
			Total efficiency & 6.0\%(5.4\%)  &  $10^{-7}$  &   $10^{-7}$   &  0.019\%  &  $10^{-7}$   &  $10^{-7}$   & 0.028\% \\
			\hline
	\end{tabular}}
\end{table}
\begin{figure}[!htb]
	\begin{center}
		\includegraphics[width=0.5\linewidth]{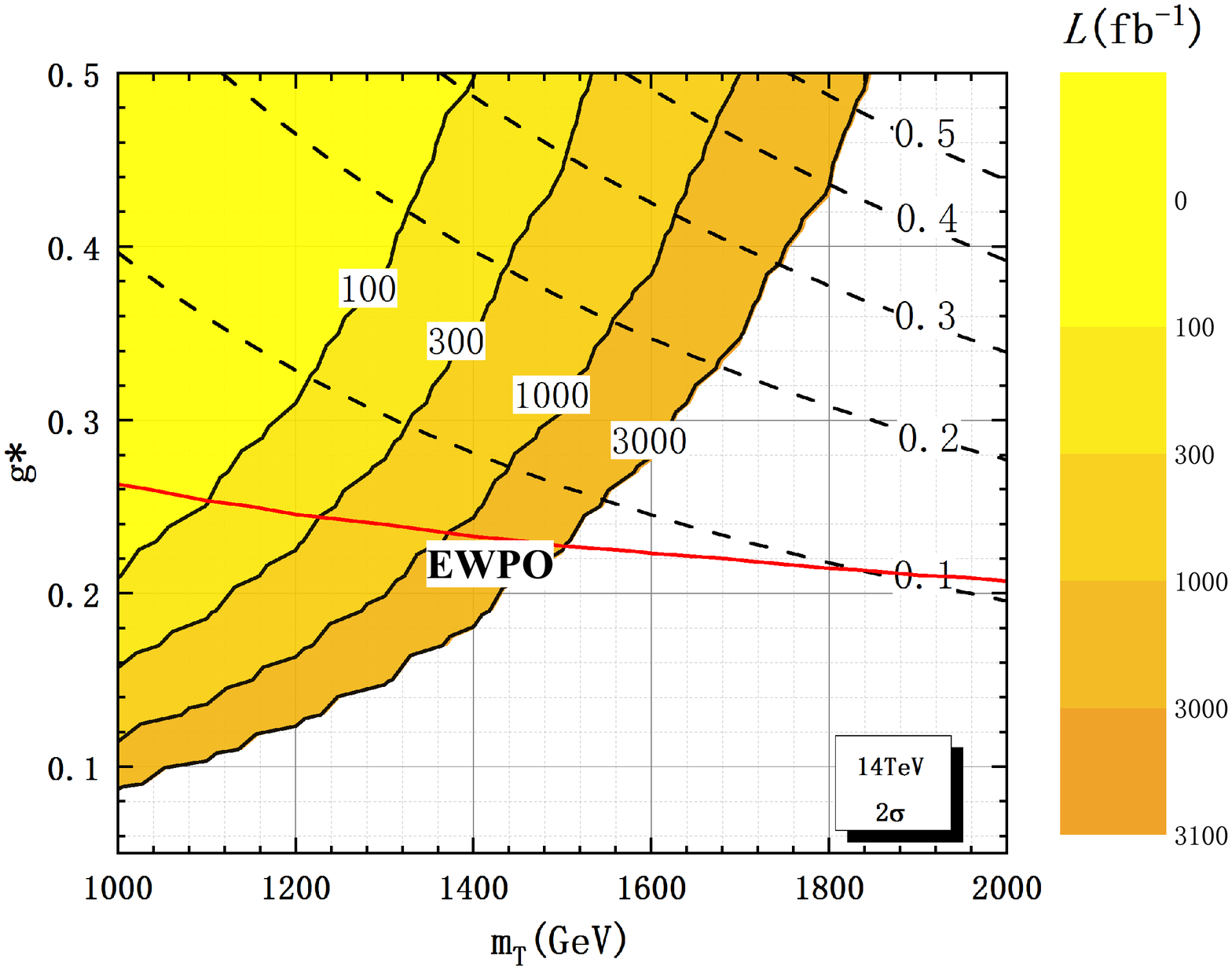}\hspace{-0.5cm}
	\includegraphics[width=0.5\linewidth]{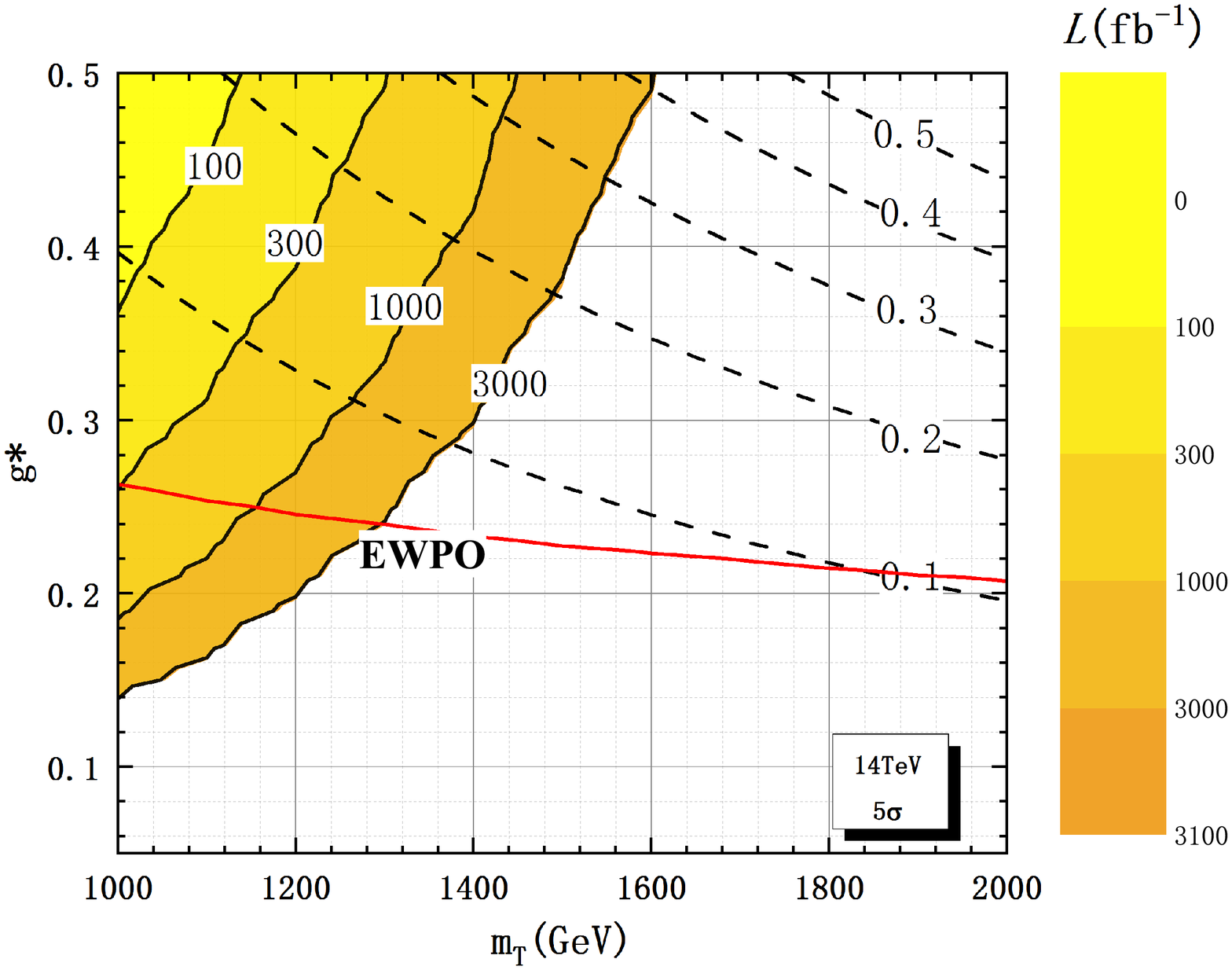}
		\caption{2$\sigma$ (left) and 5$\sigma$ (right) contour plots for the signal at $\sqrt{s}$ = 14~TeV in the $ g^*-m_T$ plane. The solid lines indicate the integrated luminosities, the dashed lines indicate the $\Gamma_T/m_T$, and the red solid lines represent the limit from EWPO at the $2\sigma$ level.}
		\label{fig:14TeV25sigma_limit}
	\end{center}
\end{figure}

For the 14 TeV LHC, we show the 2$\sigma$ exclusion (corresponding to $SS=2$) and 5$\sigma$ discovery (corresponding to $SS=5$) capabilities in the $ g^{*}-m_T $ plane in Fig.\ref{fig:14TeV25sigma_limit}, where the limit from EWPO at the $2\sigma$ level is also displayed. If the EWPO limit is not taken into account, the VLT can be excluded in the correlated regions of $ g^* $  $ \in $  [0.16,0.50] with $ m_T $ $ \in $ $[1000~\text{GeV}, 1530~\text{GeV}]$ corresponding to 300 fb$^{-1}$. For the HL-LHC with 3000 fb$^{-1}$, the excluded correlated regions can be expanded to $ g^* $  $ \in $  [0.09,0.50] with $ m_T $ $ \in $ $[1000~\text{GeV},1840~\text{GeV}]$, and the discovered correlated regions can be expanded to $ g^* $  $ \in $  [0.14,0.50] and $ m_T $ $ \in $ $[1000~\text{GeV},1600~\text{GeV}]$. 

It is worth noting that these cross sections are calculated using the narrow-width approximation (NWA). Since the widths of VLT may be large and not negligible, we also display the width-to-mass ratios $\Gamma_T$/$m_T$ in Fig.\ref{fig:14TeV25sigma_limit}. If the search at the HL-LHC is sensitive to the $\Gamma_T$/$m_T$ ranging from narrow up to 30\%, the excluded (discovered) parameter
space will be reduced to $ g^* $  $ \in $  [0.09,0.39] ([0.14,0.44]) with $ m_T $ $ \in $ $[1000~\text{GeV},1750~\text{GeV}]([1000~\text{GeV},1550~\text{GeV}])$. If the EWPO limit is considered, the excluded (discovered) parameter
space will be further reduced to $ g^* $  $ \in $  [0.09,0.23] ([0.14,0.24]) with $ m_T $ $ \in $ $[1000~\text{GeV},1500~\text{GeV}]([1000~\text{GeV},1300~\text{GeV}])$.

\subsection { $\sqrt{s}$ = 27~TeV}

\begin{figure}[!htb]
	\begin{center}
	\includegraphics[width=0.5\linewidth]{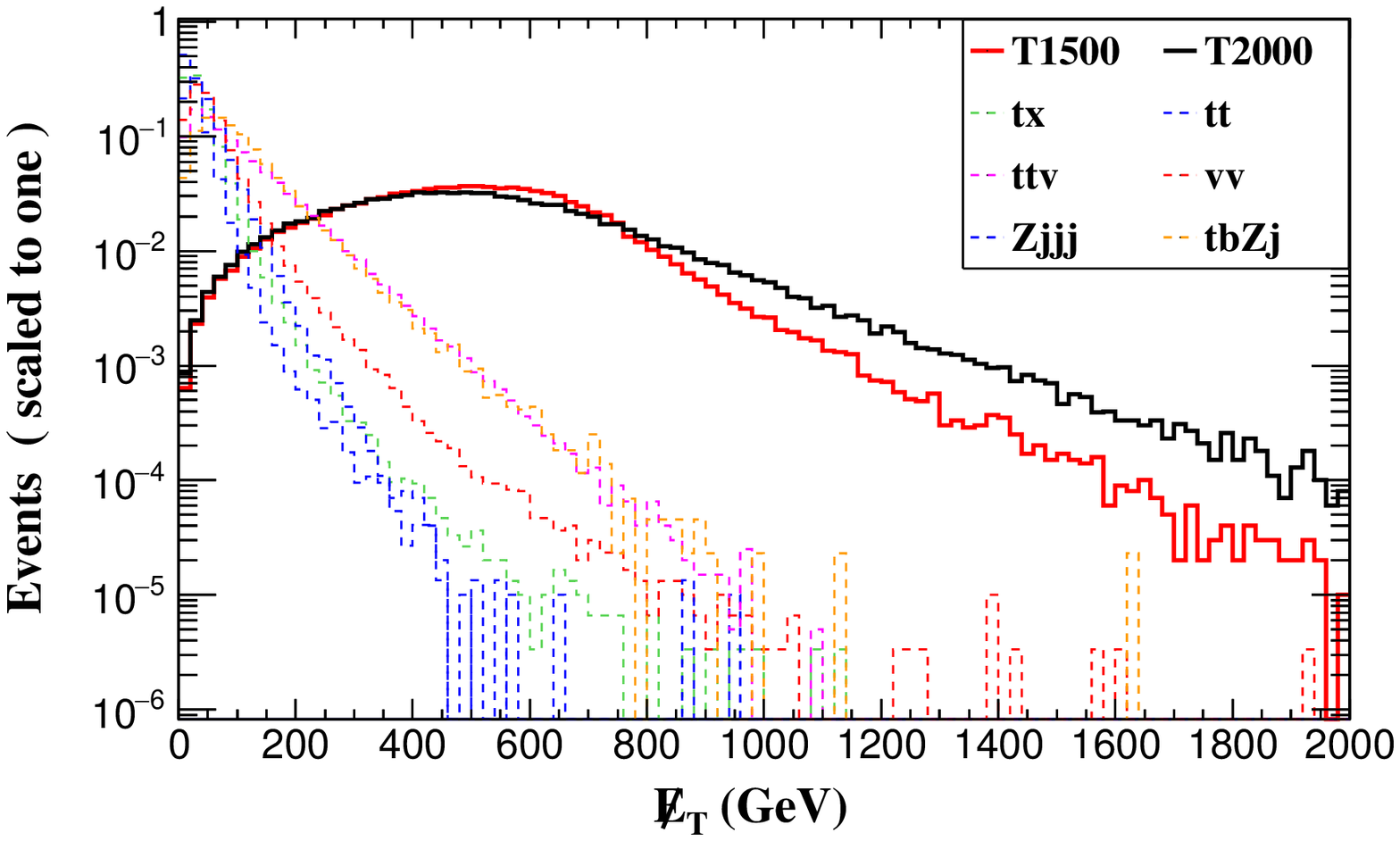}\hspace{-0.5cm}
	\includegraphics[width=0.5\linewidth]{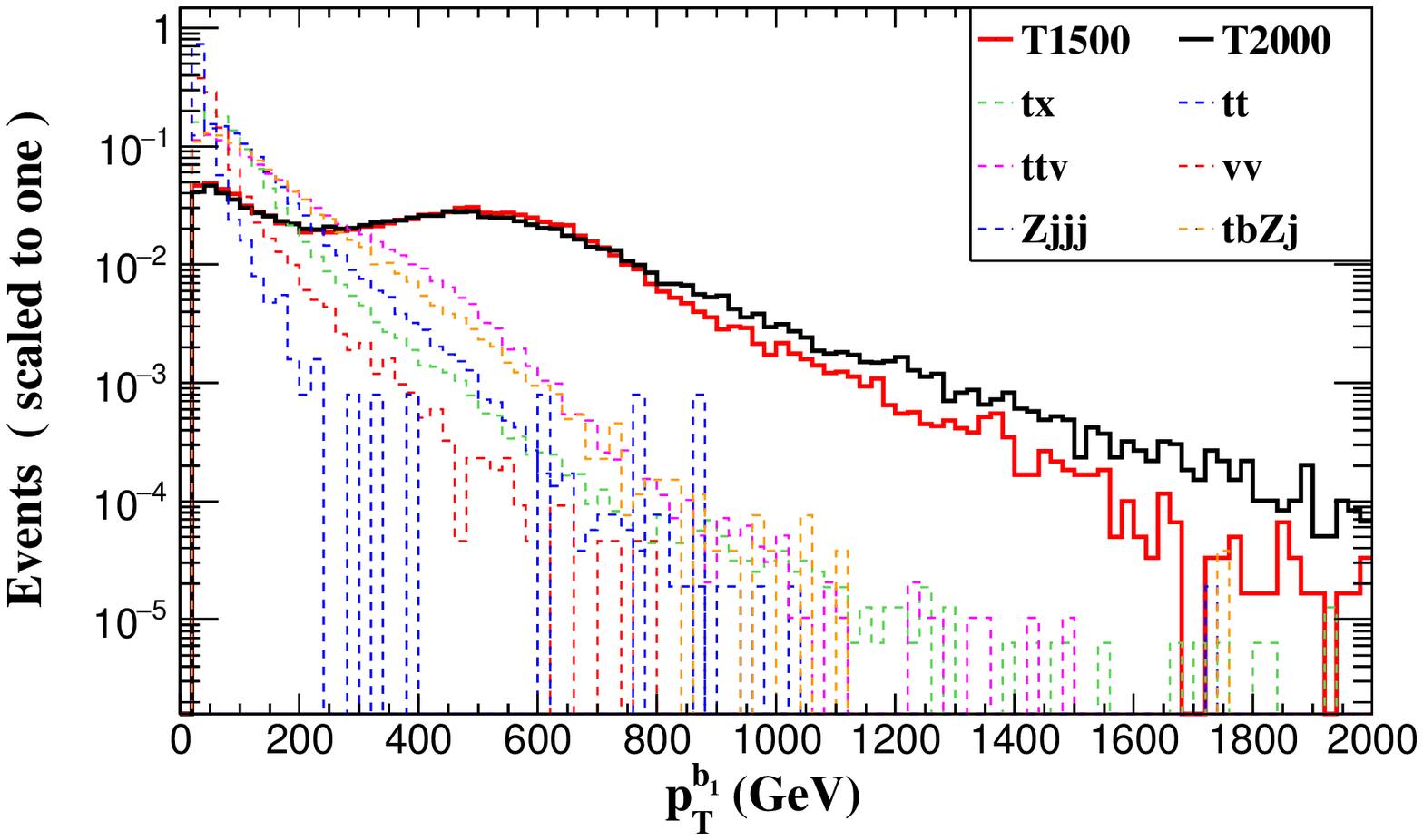}\\
	\includegraphics[width=0.5\linewidth]{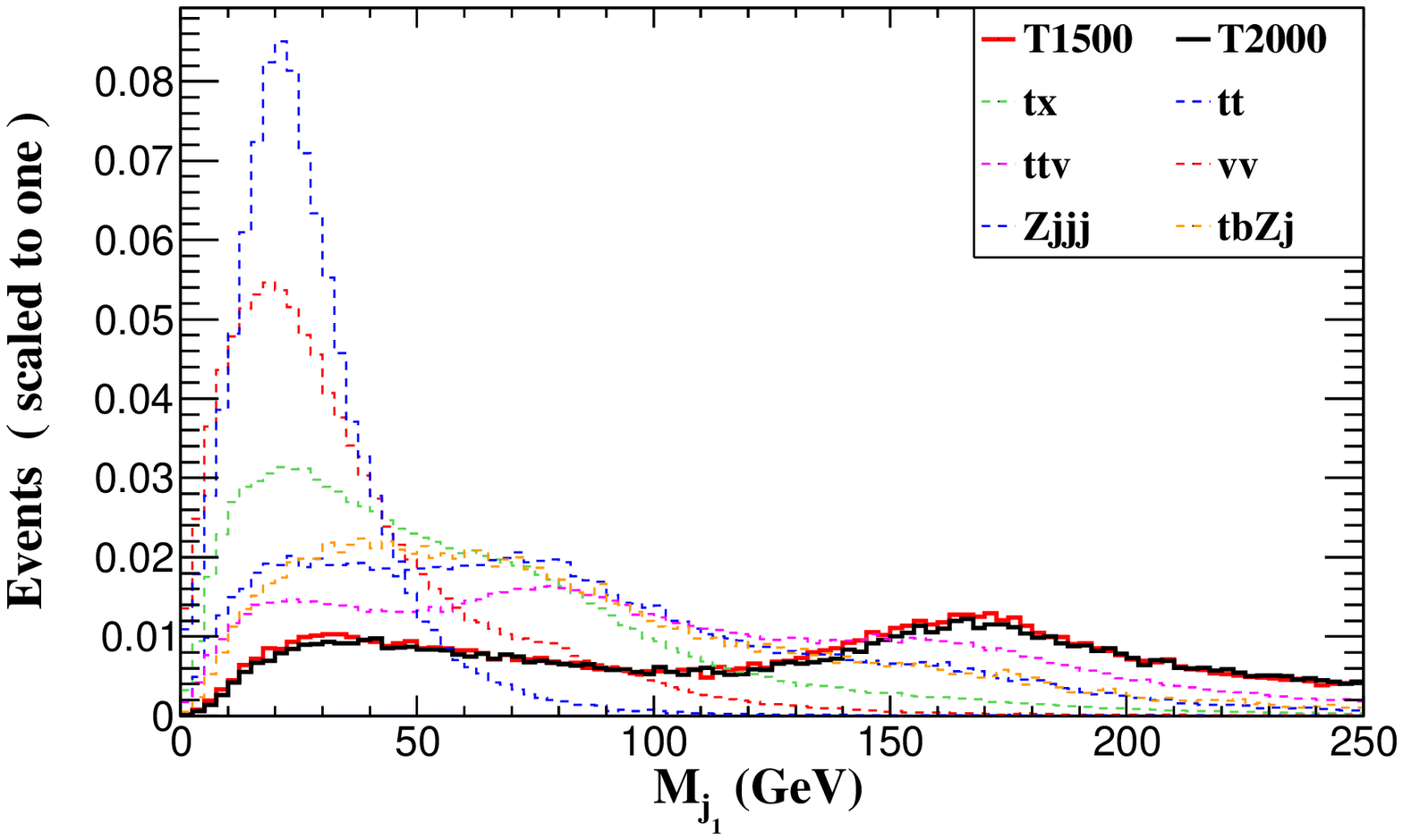}\\
		\caption{Normalized distributions of $\slashed{E}_T$, $p_T^{b_1}$, and $M_{j_1}$ for the two signal benchmark points (T1500 and T2000) and backgrounds with $g^*=0.2$ at the 27~TeV HE-LHC.
		}
		\label{fig:sm27TeV_normalized}
	\end{center}
	
\end{figure}

We take the missing energy $\slashed{E}_T> 500$ GeV, transverse momentum $p_{T}^{b_{1}}> 100 $ GeV and jet-mass 160 GeV$<M_{j_1}<$190 GeV as the selection criteria and show the normalized distributions of the signal and backgrounds for $g^*$ = 0.2 at the 27~TeV HE-LHC in Fig.\ref{fig:sm27TeV_normalized}, where we choose $m_T$ = 1500 and 2000~GeV (labeled as T1500 and T2000) as two benchmark points. 
\begin{table}[!htb]
	\centering
	\caption{Cut flows of the signal (T1500 and T2000) and backgrounds for $g^*=0.2$ at the HE-LHC. }
	\label{tab:27TeV_cutflows}
	\resizebox{1\textwidth}{!}{
		\begin{tabular}{cccccccc}
			\hline
			Cuts&\multicolumn{1}{c}{Signal (fb)}&\multicolumn{6}{c}{Backgrounds (fb)}\\
			\cline{3-8}
			& T1500(T2000)     &   $tX$    &   $t\bar{t}$  &   $ttV$  &    $VV$   &  $Zjjj$  &   $tbZj$  \\ 
			\hline
			$\sigma$ (Before cut)    & 1.19(0.24)    & 309807 & 915210  & 320      & 25000     & 690100   & 76.63  \\
			\hline
			Trigger                  & $6.27\times10^{-1}$($1.26\times10^{-1}$)  & $9.88\times10^{4}$    & $2.65\times10^{5}$  &  $8.42\times10^{1}$  &  $8.0\times10^{2}$   & $6.07\times10^{3}$   & $4.02\times10^{1}$     
			\\
			$\met > 500$  GeV        & $3.04\times10^{-1}$($6.14\times10^{-2}$) & $7.21$  &  $9.15\times10^{-2}$   &  $3.87\times10^{-1}$  &  $1.68$  &  $9.11$  & $1.81\times10^{-1}$   
			\\
			$p_{T}^{b_{1}} > 100$ GeV& $2.56\times10^{-1}$($5.31\times10^{-2}$) & $7.21$       &  $9.15\times10^{-2}$    &  $2.97\times10^{-1}$ &  $5.88\times10^{-1}$      &  $9.11$ & $1.49\times10^{-1}$
			\\
			160 GeV$< M_{j_{1}} < 190$GeV& $4.79\times10^{-2}$($8.44\times10^{-3}$) & $3.10\times10^{-2}$       &  $9.15\times10^{-2}$   &  $5.44\times10^{-2}$ &  $2.5\times10^{-3}$    &  $6.90\times10^{-2}$  & $1.38\times10^{-2}$    
			\\
			\hline
			Total efficiency         &  4.03\%(3.5\%)  &  $10^{-7}$     &  $10^{-7}$    &  0.017\%  &  $10^{-7}$    &  $10^{-7}$   &  0.018\%   
			\\
			\hline
	\end{tabular}}
\end{table}

\begin{figure}[!htb]
	\begin{center}
		\includegraphics[width=0.5\linewidth]{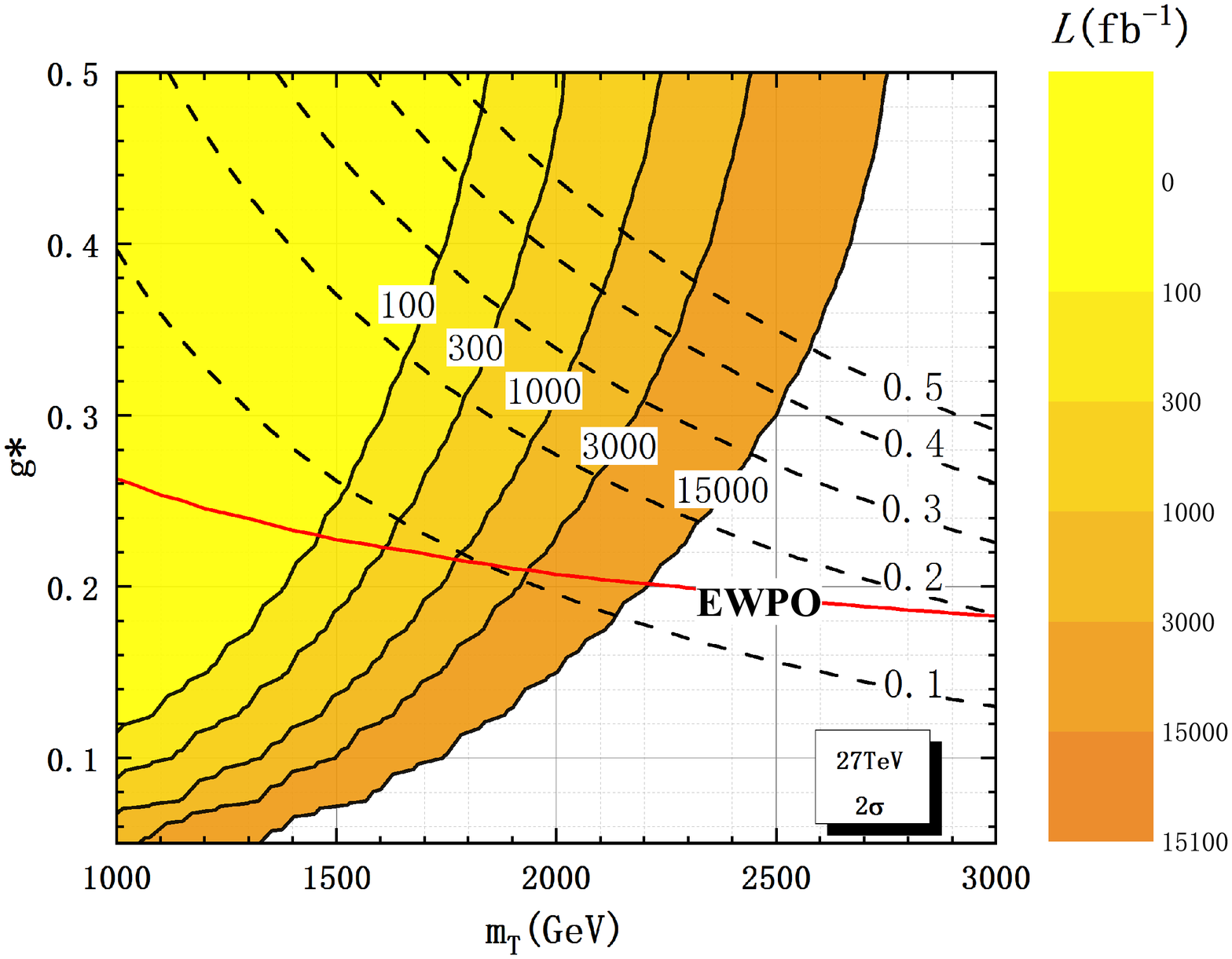}\hspace{-0.5cm}
		\includegraphics[width=0.5\linewidth]{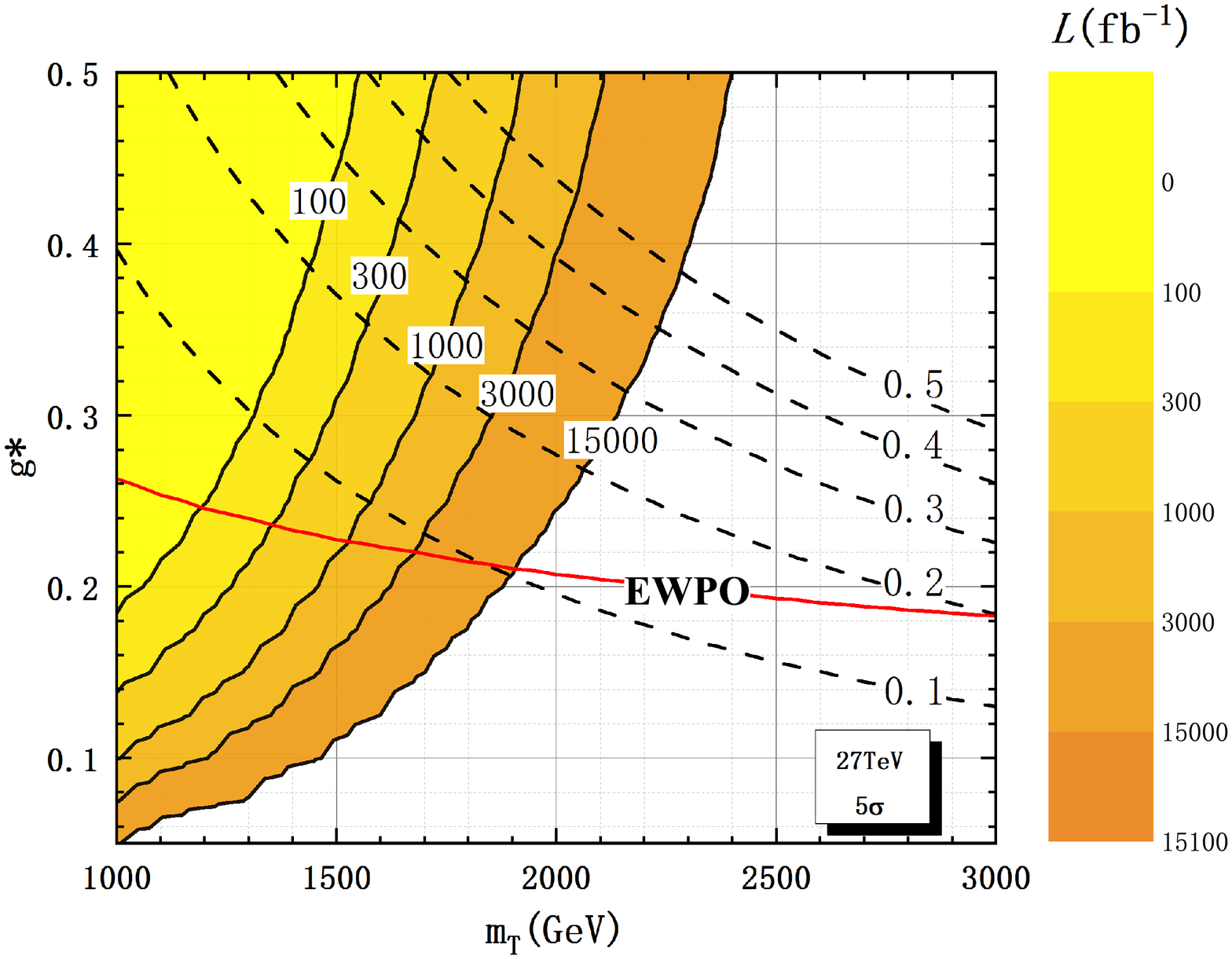}
		\caption{ Same as Fig.\ref{fig:14TeV25sigma_limit}, but for $\sqrt{s}$ = 27~TeV. }
		\label{fig:27TeV25sigma_limit}
	\end{center}
	
\end{figure}

The cut flows of the signal and the backgrounds are summarized in Table.\ref{tab:27TeV_cutflows}. We can see that the total cut efficiency of the signal can reach 4.03\% (3.5\%) for the benchmark point T1500 (T2000), while the backgrounds can be suppressed effectively. Similar to the LHC, we scan the parameter space $g^*\in[0.05, 0.5]$ and 
$m_{T}\in[1000\text{GeV}, 3000\text{GeV}]$ and impose the average value 3.77\% of these two signal efficiencies to the entire parameter space.
The exclusion and discovery capabilities
in the $ g^*-m_T $ plane at the HE-LHC are shown in Fig.\ref{fig:27TeV25sigma_limit}. Compared to the data of the HL-LHC,
the excluded (discovered) correlated regions of the VLT at HE-LHC can be expanded to  $ g^* $  $ \in $  [0.05,0.50]([0.08,0.50]) with $ m_T $ $ \in $ $[1000~\text{GeV},2440~\text{GeV}]([1000~\text{GeV},2100~\text{GeV}])$ corresponding to 3000 fb$^{-1} $. For the HE-LHC with 15 ab$^{-1}$, the correlated regions $ g^* $  $ \in $  [0.05,0.50] with $ m_T $ $ \in $ $[1300~\text{GeV},2750~\text{GeV}]([1000~\text{GeV},2400~\text{GeV}])$ can be excluded (discovered). If the limit $\Gamma_T/m_T<$ 30\% is considered, the excluded (discovered) regions will be reduced to $ g^* $  $ \in $  [0.05,0.28] ([0.05,0.31]) with $ m_T $ $ \in $ $[1300~\text{GeV},2440~\text{GeV}]([1000~\text{GeV},2160~\text{GeV}])$ corresponding to 15 ab$^{-1} $. If the EWPO limit is considered, the excluded (discovered) regions will be further reduced to $ g^* $  $ \in $  [0.05,0.20] ([0.05,0.21]) with $ m_T $ $ \in $ $[1300~\text{GeV},2210~\text{GeV}]([1000~\text{GeV},1900~\text{GeV}])$. 

\subsection { $\sqrt{s}$ = 100~TeV}

\begin{figure}
	\includegraphics[width=0.5\linewidth]{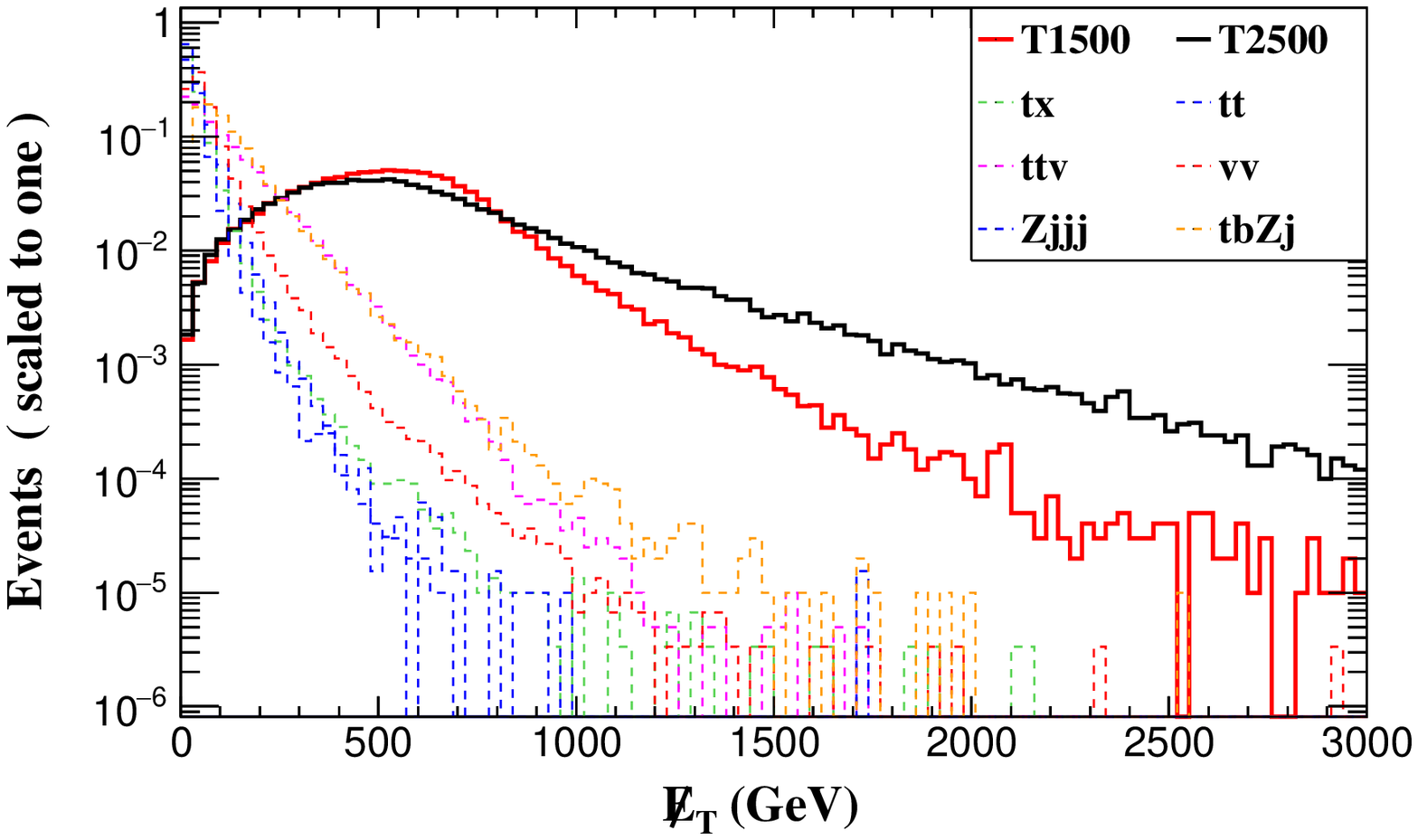}\hspace{-0.5cm}
	\includegraphics[width=0.5\linewidth]{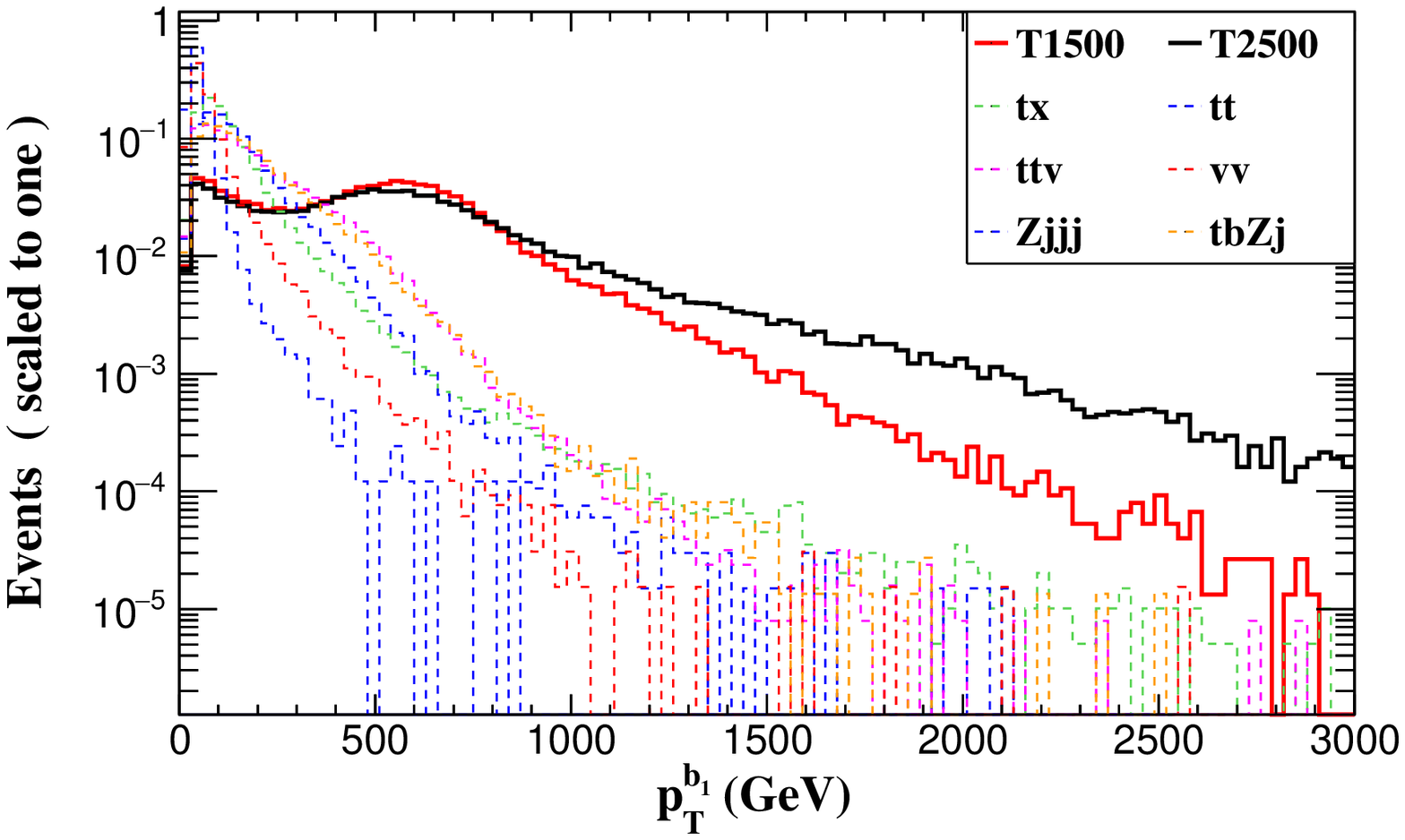}\\
	\includegraphics[width=0.5\linewidth]{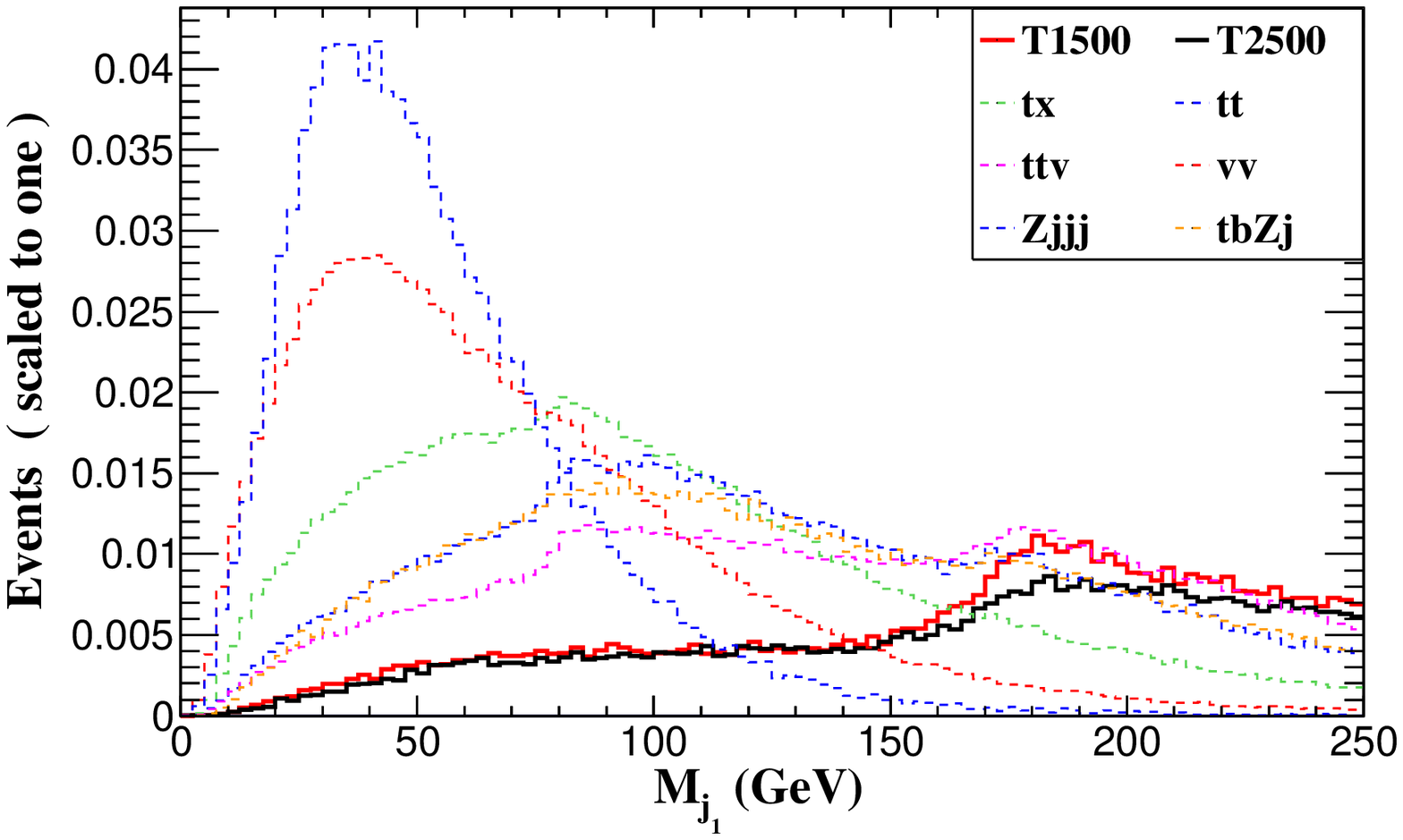}\\
	\caption{Normalized distributions of $\slashed{E}_T$, $p_T^{b_1}$, and $M_{j_1}$ for the two signal benchmark points (T1500 and T2500) and backgrounds with $g^*=0.2$ at the 100~TeV FCC-hh.
	}
	\label{fig:sm100TeV_normalized}
\end{figure}

We take the missing energy $\slashed{E}_T>650$ GeV, transverse momentum $p_{T}^{b_{1}}> 500$ GeV, and jet-mass 160 GeV$<M_{j_1}<$190 GeV as the selection criteria and show these normalized distributions of the signal and backgrounds for $g^*$ = 0.2 at 100~TeV FCC-hh in Fig.\ref{fig:sm100TeV_normalized}, where $m_T$ = 1500 and 2500~GeV (labeled as T1500 and T2500) are chosen as two benchmark points. The cut flows of the signal and backgrounds are summarized in Table.\ref{tab:100TeV_cutflows}. We can see that the total cut efficiency of the signal can reach 1.4\% (1.3\%) for the benchmark point T1500 (T2500). Similarly, we scan the parameter space $g^*\in[0.05, 0.5]$ and $m_{T}\in[1000\text{GeV}, 4000\text{GeV}]$ and impose the average value 1.35\% of these two signal efficiencies to the entire parameter space.

\begin{table}[!htb]
	\centering
	\caption{Cut flows of the signal (T1500 and T2500) and backgrounds for $g^*=0.2$ at the FCC-hh. }
	\label{tab:100TeV_cutflows}
	\resizebox{1\textwidth}{!}{
		\begin{tabular}{cccccccc}
			\hline
			Cuts&\multicolumn{1}{c}{Signal (fb)}&\multicolumn{6}{c}{Backgrounds (fb)}\\
			\cline{3-8}
			& T1500(T2500)    &   $tX$    &   $t\bar{t}$   &   $ttV$  &    $VV$   &  $Zjjj$  &   $tbZj$ \\ 
			\hline
			$\sigma$ (Before cut)    &  21.36(1.96)    & 1879973   & 9510025  &  3720  & 89138   & 5701000  &  779.3  \\
			\hline
			Trigger  & $1.24\times10^{1}$(1.12)    & $6.94\times10^{5}$    & $3.12\times10^{6}$  &  $1.15\times10^{3}$  &  $9.45\times10^{3}$  & $5.02\times10^{5}$    &  $4.40\times10^{2}$     
			\\
			$\met > 650$  GeV       & $3.69$($4.22\times10^{-1}$)    & $1.04\times10^{2}$     &  $1.87\times10^{2}$   &  $3.33$   & $1.32\times10^{1}$   &  $1.76\times10^{2}$    &  $1.76$   
			\\
			$p_{T}^{b_{1}} > 150$ GeV      & $3.25$($3.79\times10^{-1}$)   & $9.76\times10^{1}$   &  $1.87\times10^{2}$  &  $2.73$  &   $5.85$  &  $8.78\times10^{1}$ &  $1.42$  
			\\
			160 GeV$< M_{j_{1}} < 190$ GeV     & $2.99\times10^{-1}$($2.55\times10^{-2}$)  &  $1.88\times10^{-1}$    &  $9.51\times10^{-1}$   &  $4.09\times10^{-1}$  &  $2.94\times10^{-1}$    & $5.70\times10^{-1}$  &   $1.01\times10^{-1}$     
			\\
			\hline
			Total efficiency & 1.4\%(1.3\%)  &  $10^{-7}$    & $10^{-7}$   &  0.011\%  &  0.00033\%    & $10^{-7}$  &   0.013\%   \\
			\hline
	\end{tabular}}
\end{table}

\begin{figure}[!htb]
	\begin{center}
		\includegraphics[width=0.5\linewidth]{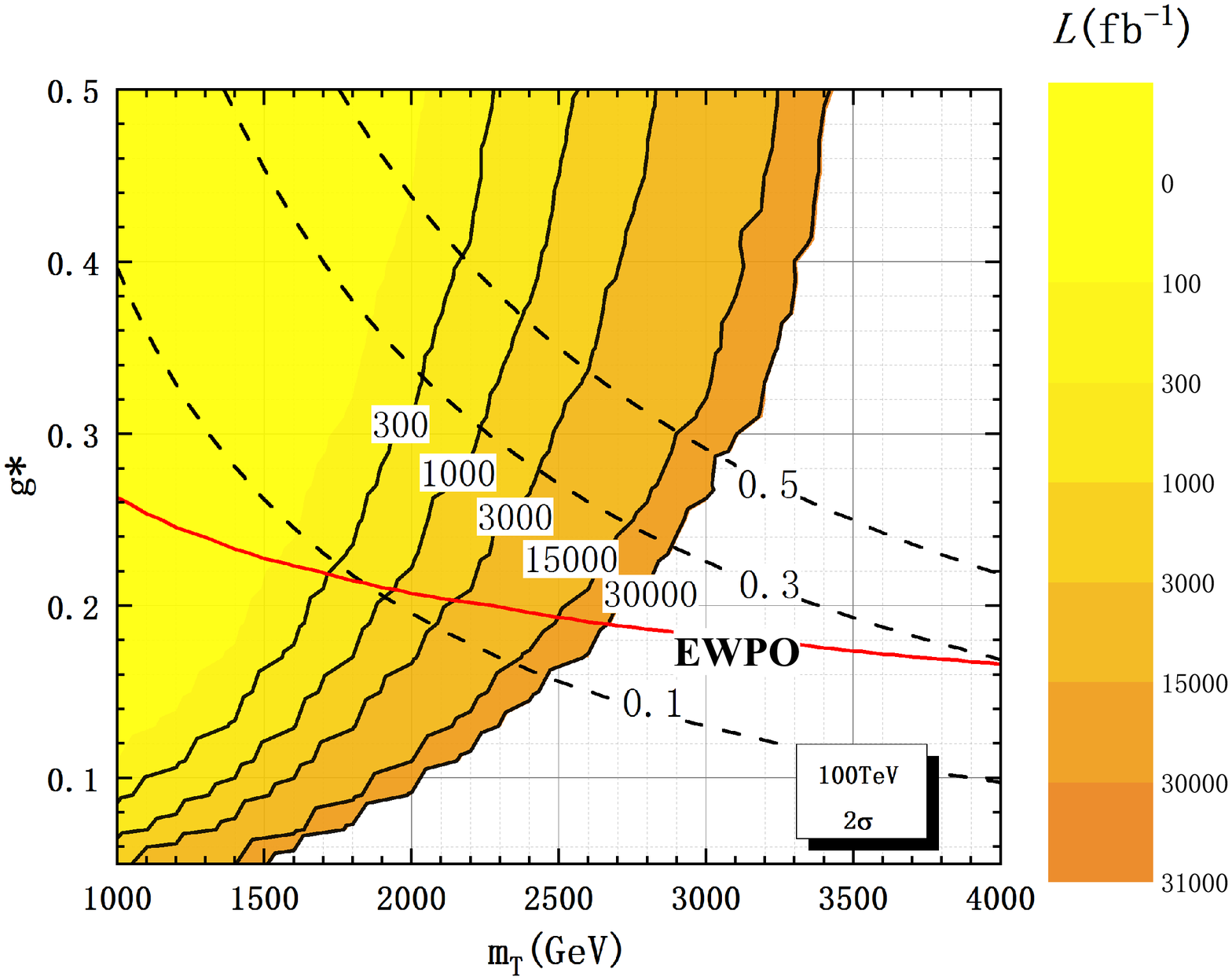}\hspace{-0.5cm}
		\includegraphics[width=0.5\linewidth]{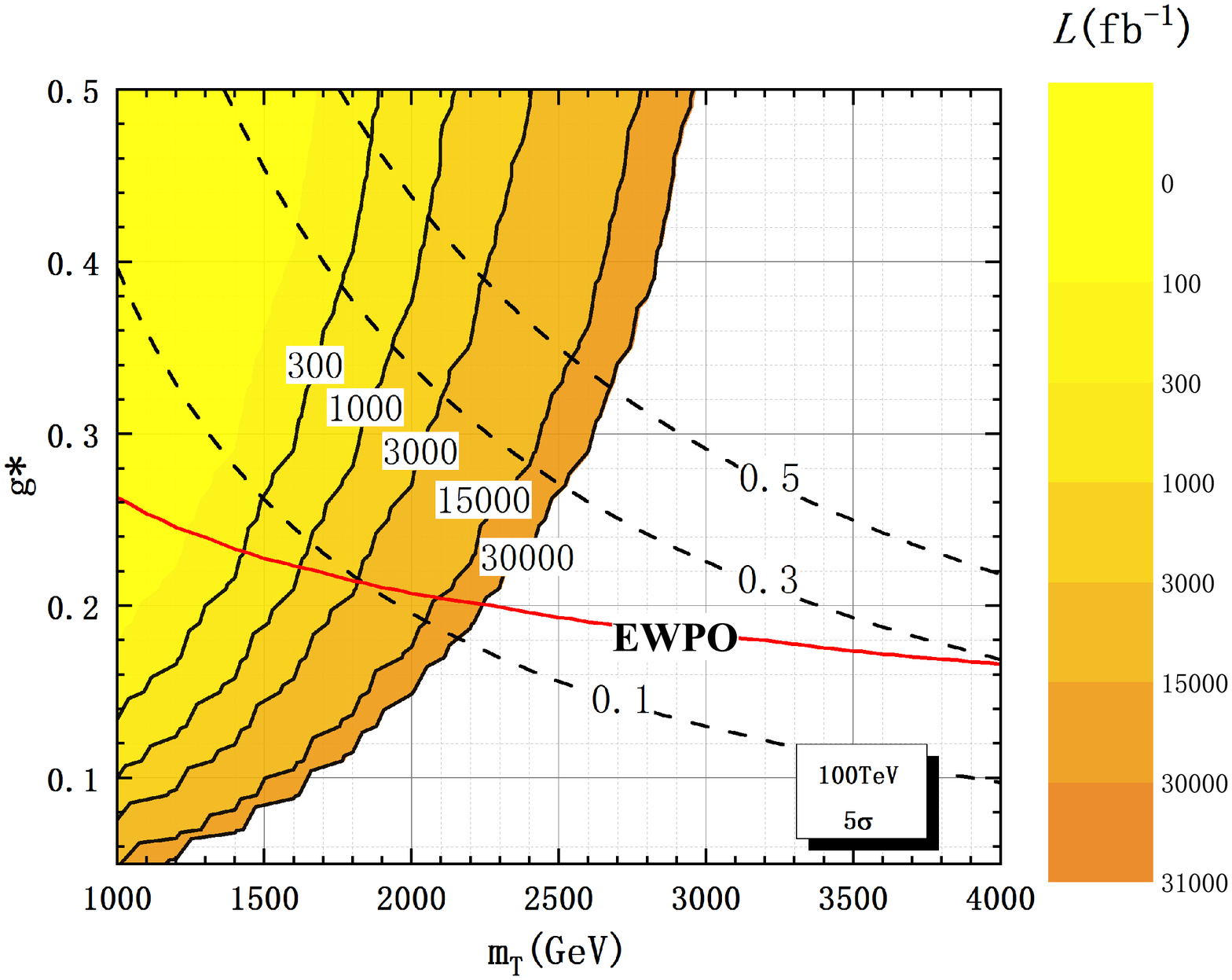}\\
		\caption{ Same as Fig.\ref{fig:14TeV25sigma_limit}, but for $\sqrt{s}$ = 100~TeV. }
		\label{fig:100TeV25sigma_limit}
	\end{center}
	
\end{figure}

The exclusion and discovery capabilities in the $g^*-m_T $ plane at $\sqrt{s}$ = 100~TeV are shown in Fig.\ref{fig:100TeV25sigma_limit}. Corresponding to 3000fb$^{-1}$ (15ab$^{-1}$), the $T$ quark can be excluded in the correlated regions of $ g^* $  $ \in $ [0.05,0.50]([0.08,0.50]) with $ m_T $ $ \in $ $ [1050~\text{GeV},2810~\text{GeV}]([1400~\text{GeV},3240~\text{GeV}])$. For the FCC-hh with 30 ab$^{-1}$, the excluded (discovered) regions can be expanded to $ g^* $  $ \in $  [0.05,0.50] with $ m_T $ $ \in $ $[2500~\text{GeV}, 3420~\text{GeV}] ([2050~\text{GeV}, 2950~\text{GeV}])$.
If the limit $\Gamma_T$/$m_T$ < 30\% is considered, the excluded (discovered) regions will be reduced to $ g^* $  $ \in $  [0.05,0.24] ([0.05,0.27]) with $ m_T $ $ \in $ $[2500~\text{GeV},2880~\text{GeV}]([2050~\text{GeV},2510~\text{GeV}])$ corresponding to 30 ab$^{-1}$. If the EWPO limit is considered, the excluded (discovered) regions will be further reduced to $ g^* $  $ \in $  [0.05,0.19] ([0.05,0.20]) with $ m_T $ $ \in $ $[2500~\text{GeV},2660~\text{GeV}]([2050~\text{GeV},2240~\text{GeV}])$.

\section{SUMMARY}
In this study, we investigate the single production of the VLT decaying into $tZ$ with $Z\to \nu\bar{\nu}$ at the HL-LHC, HE-LHC, and FCC-hh. We utilize a simplified model including a $SU(2)$ singlet $T$ with charge 2/3, and the $T$ quark couples exclusively to the third-generation
SM quarks. At this time, only the mass $m_T$ and coupling constant $g^{*}$ are the free parameters. Under the limits of LHC direct searches, we perform a detailed detector simulation for the signal and backgrounds. We summarize the exclusion and discovery capabilities on the $T$ quark at different hadron colliders with the highest designed integrated luminosity in Table.\ref{tab:2sigma5sigma}, where the results from the limits  $\Gamma_T$/$m_T$ < 30\% and EWPO are also listed.  

\begin{table}[!htb]
	\centering
	\caption{Exclusion and discovery capabilities on $T$ at different hadron colliders.}
	\label{tab:2sigma5sigma}
	\resizebox{0.9\textwidth}{!}{
		\begin{tabular}{c|c|c|c||c|c|c}
			\hline\hline
			& \multicolumn{3}{c||}{Exclusion  Capability (2$\sigma$)}                                                                                                                                                                                                                             & \multicolumn{3}{c}{Discovery Capability (5$\sigma$)}                                                                                                                                                                                                                                \\ \hline
			Colliders     & \begin{tabular}[c]{@{}c@{}}HL-LHC \\\end{tabular} & \multicolumn{1}{c|}{\begin{tabular}[c]{@{}c@{}}HE-LHC\\\end{tabular}} & \begin{tabular}[c]{@{}c@{}}FCC-hh \\\end{tabular} & \multicolumn{1}{c|}{\begin{tabular}[c]{@{}c@{}}HL-LHC\\ \end{tabular}}&
			\multicolumn{1}{c|}{\begin{tabular}[c]{@{}c@{}}HE-LHC\\ \end{tabular}}&
			\multicolumn{1}{c}{\begin{tabular}[c]{@{}c@{}}FCC-hh\\ \end{tabular}} \\ \hline
			Luminosity  &$\mathcal{L}$=3ab$^{-1}$ &$\mathcal{L}$=15ab$^{-1}$ &$\mathcal{L}$=30ab$^{-1}$  
			&$\mathcal{L}$=3ab$^{-1}$
			&$\mathcal{L}$=15ab$^{-1}$ &$\mathcal{L}$=30ab$^{-1}$ 
			\\ \hline
			\begin{tabular}[c]{@{}c@{}}$g^*$\end{tabular}  
			&{[}0.09,0.5{]}      & {[}0.05,0.5{]}      & {[}0.05,0.5{]}    &{[}0.14,0.5{]} &{[}0.05,0.5{]}&{[}0.05,0.5{]}           \\ \hline
			$m_T$(GeV)     & {[}1000,1840{]}      & {[}1300,2750{]}       & {[}2500,3420{]}          & {[}1000,1600{]} &{[}1000,2400{]}&{[}2050,2950{]}        \\  \hline \hline
			\begin{tabular}[c]{@{}c@{}}$g^*$$(\frac{\Gamma_T}{m_T}$< 30\%)\end{tabular}  
			&{[}0.09,0.39{]}      & {[}0.05,0.28{]}      & {[}0.05,0.24{]}    &{[}0.14,0.44{]} &{[}0.05,0.31{]}&{[}0.05,0.27{]}         \\ \hline
			$m_T$(GeV)     & {[}1000,1750{]}      & {[}1300,2440{]}       & {[}2500,2880{]}          & {[}1000,1550{]} &{[}1000,2160{]}&{[}2050,2510{]}    \\   \hline \hline
			\begin{tabular}[c]{@{}c@{}}$g^*$(EWPO)\end{tabular}  
			&{[}0.09,0.23{]}      & {[}0.05,0.20{]}      & {[}0.05,0.19{]}    &{[}0.14,0.24{]} &{[}0.05,0.21{]}&{[}0.05,0.20{]}         \\ \hline
			$m_T$(GeV)     & {[}1000,1500{]}      & {[}1300,2210{]}       & {[}2500,2660{]}          & {[}1000,1300{]} &{[}1000,1900{]}&{[}2050,2240{]}     \\  \hline \hline
	\end{tabular}}
	
\end{table}
We can see that the exclusion and discovery capabilities on the $T$ quark are enhanced evidently with the increase in the collision energy. If we consider the NWA and EWPO limits, the exclusion and discovery regions will be reduced to some extent. We expect these results to provide a meaningful reference for the search for such a singlet VLT quark at future hadron colliders.

\section*{Acknowledgement}
This work is supported by the National Natural Science Foundation of China (NNSFC) under Grants No. 11705048, the National Research Project Cultivation Foundation of Henan Normal University under Grant Nos. 2020PL16, 2021PL10, the Startup
Foundation for Doctors of Henan Normal University under Grant No. qd18115, and also powered by the High Performance Computing Center of Henan Normal University.

\end{document}